# High spin polarization and large spin splitting in equiatomic quaternary CoFeCrAl Heusler alloy

Lakhan Bainsla,[1] A. I. Mallick,[1] A. A. Coelho,[2] A. K. Nigam,[3] B. S. D. Ch. S. Varaprasad,[4] Y. K. Takahashi,[4] Aftab Alam,[1] K. G. Suresh,[1,#] and K. Hono[4]

[1]Department of Physics, Indian Institute of Technology Bombay, Mumbai 400076, India
[2]Instituto de Física "Gleb Wataghin", Universidade Estadual de Campinas-UNICAMP, SP 6165, Campinas 13 083 970, Sao Paulo, Brazil
[3]Department of Condensed Matter and Materials Science, Tata Institute of Fundamental Research, Mumbai 4000052
[4]Magnetic Materials unit, National Institute for Materials Science, Tsukuba 305-0047, Japan

**Abstract**

In this paper, we investigate CoFeCrAl alloy by means of various experimental techniques and *ab-initio* calculations to look for half-metallic nature. The alloy is found to exist in the cubic Heusler structure, with presence of B2 ordering. Saturation magnetization ($M_S$) value of about 2 $\mu_B$/f.u. is observed at 8 K under ambient pressure, which is in good agreement with the Slater-Pauling rule. $M_S$ values are found to be independent of pressure, which is a prerequisite for half-metals. The *ab-initio* electronic structure calculations predict half-metallic nature for the alloy with a spin slitting energy of 0.31 eV. Importantly, this system shows a high current spin polarization value of 0.67 ± 0.02, as deduced from the point contact Andreev reflection (PCAR) measurements. Linear dependence of electrical resistivity with temperature indicates the possibility of reasonably high spin polarization at elevated temperatures (~150 K) as well. All these suggest that CoFeCrAl is a promising material for the spintronic devices.

PACS: 81.05.Bx; 75.47.Np; 85.75.-d; 31.15.A-

Key words: Heusler alloys, half metallic ferromagnet, spintronics, *ab-initio* calculation, PCAR

Due to their high spin polarization, half-metallic ferromagnetic (HMF) materials have potential use in spintronic devices. Recently Heusler alloys attracted a lot of attention due to their theoretically predicted half-metallic nature.[1] Among these, Co based alloys have a special place due to their experimentally observed high spin polarization (P) and high Curie temperatures.[2-6] HMF have applications as spin polarized current sources for current-perpendicular-to-plane giant magnetoresistive (CPP-GMR) devices,[7-8] magnetic tunneling junctions,[9] lateral spin valves,[10] and spin injectors to semiconductors.[11] High values of P were observed at low temperatures for certain Co based alloys such as $Co_2Fe(Ga_{0.5}Ge_{0.5})$,[3] and $Co_2(Fe_{0.4}Mn_{0.6})Si$,[9] but a strong degradation in the spin transport properties is observed at room temperature for the devices because the value of P decreases drastically with temperature. Therefore, further exploration of half-metallic Heusler alloys is strongly desired.

The full Heusler alloys $X_2YZ$ (where X and Y are transition and Z is a main group element) exist in $L2_1$ crystal structure with four interpenetrating sub-lattices. If each lattice site is occupied by a different element, a quaternary Heusler structure with changed symmetry is obtained, which is named as LiMgPdSn type.[12] Based on the occupation of different lattice sites, there are three sub-classes of LiMgPdSn type structure. To the best of our knowledge, equiatomic quaternary Heusler alloys (with 1:1:1:1 stoichiometry) is explored only very little, but some of these alloys are found to show interesting properties such as high spin polarization and spin gapless semiconductor behavior.[4,13-16] The electrical resistivity in equiatomic quaternary Heusler alloys is supposed to be less than that observed in pseudo-ternary alloys such as $X_2Y_{1-x}Y'_xZ$, because random distribution of Y and Y' leads to an additional scattering. In these latter alloys, the spin diffusion length becomes very short, only in the order of a few nm.[17]

The equiatomic quaternary Heusler alloy CoFeCrAl (CFCA) can be regarded as the combination of $Co_2FeAl$ and $Co_2CrAl$. CFCA has been studied earlier,[18] but information about the transport (resistivity and spin polarization) studies, which are very important and conclusive in half-metallic point of view, is lacking. In the present work, CFCA is studied by *ab-initio* calculations as well as investigated experimentally to look for half-metallic properties. These studies show that the alloy is half-metallic in nature with a high value of current spin polarization at 4.2 K. An estimate of spin polarization is extremely useful in spin transport based applications.

Polycrystalline bulk alloy (CFCA) was prepared using arc melting technique. Ti ingot was melted before melting the sample to avoid any oxidation. The sample was melted several times to ensure the homogeneity and to further increase the homogeneity, it was sealed in evacuated quartz tube under high vacuum and heat treated at 1073 K for 14 days, followed by cold water quenching. Structural analysis of the sample was done by room temperature x-ray diffraction (XRD) measurements in a Philips diffractometer with Cu-K$_\alpha$ radiation. Magnetization isotherms at 8 K under different pressures have been measured with the help of a SQUID magnetometer. Spin polarization measurements were done using the point contact Andreev reflection spectroscopy (PCAR),[19] where the conductance curves were measured at the point contact between a superconductor and the sample. To make the superconducting point contacts, sharp Nb tips were prepared by electrochemical polishing. Spin polarization of the conduction electrons were obtained by fitting the conductance curves to modified BTK model,[20] using spin polarization (P), superconducting band gap (Δ) and interfacial scattering parameter (Z) as variables.

Self-consistent band structure calculations were performed with spin-polarized density functional theory (DFT) using Vienna ab-initio simulation package (VASP).[21] The exchange-correlation functionals were treated under the generalized gradient approximation (GGA) with the projected-augmented wave (PAW) basis.[22] A Monkhorst-pack Brillouin zone integration with 16x16x16 k-mesh (resulting in 145 k-points in the irreducible wedge of Brillouin zone) was used for the calculations.[23] We have used a high plane wave cut-off of 340 eV with the energy convergence criteria of 0.1 meV/cell.

Quaternary Heusler alloy CoFeCrAl crystallizes in a LiMgPdSn (Y-type) structure with space group F-43m (# 216). Out of different inequivalent super-structures (based on the occupation of Wyckoff sites by different constituent elements) for Y-type structure, we found from total energy calculations that the super-structure with Co @ 4a(0,0,0), Fe @ 4b(1/2,1/2,1/2), Cr @ 4c(1/4,1/4,1/4), Al @ 4d(3/4,3/4,3/4) to be the most stable one.

The spin resolved band structure and density of states (DOS) of CoFeCrAl calculated for the most stable state with experimental lattice parameter ($a_{expt}$ = 5.75 Å) is shown Fig. 1. The band structure in the minority-spin state shows an open band gap (~ 0.31 eV) across the Fermi level, whereas for the majority-spin state, the occurrence of some valence bands cross the Fermi level gives rise to finite density of states at $E_F$. These electronic behaviors suggest it to be half-metallic in nature, which is confirmed by our experiment results (high current spin polarization value). Half-metallicity with a band gap of 0.31 eV obtained from our electronic structure calculations agrees with that calculated by Gao *et al*.[24] The calculated total magnetic moment ($\mu_{total}$) of CFCA is 1.97 $\mu_B$, which is in agreement with the Slater-Pauling rule.[25] It may be noted that for a half-metal, $\mu_{total} = Z_t - 24$, where $Z_t$ is the total number of valence electrons per formula unit (which is 26 for CFCA).

The room temperature powder x-ray diffraction pattern of the alloy is shown in the Fig. 2. It is found to exist in the cubic Heusler structure; all the fundamental peaks [(220), (400) and (422)] are present in the XRD. Superlattice reflection (200) is also found to be present, but superlattice (111) peak is absent, which indicates the presence of B2 type ordering in the structure. The degree of the long range B2 ordering ($S_{B2}$), can be roughly estimated from the intensity ratio of (200) peak to (400) fundamental peak.[26] $S_{B2}$ value of 0.89 is obtained, which indicates the presence of a highly ordered B2 structure. The B2 ordering was found to be necessary condition to obtain a half-metallic electronic structure.[27] The obtained lattice parameter of 5.75 Å is in good agreement with the earlier report.[18]

Isothermal magnetization curves were obtained at 8 K as shown in the Fig. 3. The saturation magnetization ($M_S$) value of $\approx 2.0$ $\mu_B$/f.u. is observed under ambient pressure, and it is found to be independent of applied pressure, as shown in the Fig. 3. Such a behavior, seen also in $Co_2TiGa$,[28] is the characteristic feature of a half-metallic ferromagnet. As mentioned earlier, the observed $M_S$ value is in good agreement with that calculated by the Slater-Pauling rule ($M_S = 2.0$ $\mu_B$/f.u.).[25] The effect of lattice compression on the half-metallicity in certain Heusler alloys has been studied earlier from *ab*-initio calculation by Picozzi et al., and predicted that the half-metallicity is preserved under 3-5 % variation of lattice parameter with increased value of half-metallic band gap.[29] The Curie temperature ($T_C$) of the alloy is found to be more than 400 K as there was no magnetic transition observed in the *M vs. T* curves up to 400 K. This is consistent with the report of $T_C = 460$ K by Luo et *al.* [18]

The conductance curves at 4.2 K were recorded (Fig. 4) at the ferromagnetic/superconductor point contact by PCAR technique using Nb as a superconducting tip. The spin polarization was obtained by fitting the normalized conductance curves to the modified BTK model using P, superconducting band gap ($\Delta$) and Z as variable parameters.[20] We assumed $\Delta = \Delta_1 = \Delta_2$ due to the absence of proximity effect in PCAR spectra, and the fitted parameters for the best fitting are shown in the figures. The P obtained by PCAR is the transport spin polarization and is a very important parameter from the application point of view. The transport spin polarization can be equal to the actual spin polarization ($P_a$) as $P_a = \{N_\uparrow(E_F) - N_\downarrow(E_F)\}/\{N_\uparrow(E_F) + N_\downarrow(E_F)\}$; where $N_{\uparrow(\downarrow)}(E_F)$ is the DOS at the Fermi level for spin up (down) electrons only when the Fermi energy for all the carriers is equal.[30] Depending on the value of Z, the shape of the conductance curves changes near $\Delta$ as the curves become flat near $\Delta$ for low Z values. The obtained values of $\Delta$ for the best fitting are found to be less than the bulk superconducting band gap of Nb (1.5 meV), which is attributed to the multiple contacts at the interface.[31] Such a difference has been observed earlier for CoFeMnGe,[4] $Co_2FeGa_{0.5}Ge_{0.5}$[3] and a few other Heusler alloys. The intrinsic value of current spin polarization can be achieved by recording the conductance curves at a ballistic point contact (Z = 0) with the sample, but in our measurements the lowest value of Z was found to be equal to 0.13. Therefore, the intrinsic value of P can be deduced by extrapolating the P *vs.* Z curve to Z = 0. The current spin polarization thus obtained is $0.67 \pm 0.02$, which is comparable to the P value obtained for some highly spin polarized Heusler alloys such as CoFeMnGe with $P = 0.70 \pm 0.01$,[4] $Co_2Fe(Ga_{0.5}Ge_{0.5})$ with $P = 0.69 \pm 0.02$.[3]

Electrical resistivity ($\rho$) of the alloy was measured using the conventional four probe technique in the temperature range of 5-300 K as shown in Fig. 5. $\rho(T)$ follows a linear behavior with temperature up to 150 K and is described very well by the relation $\rho(T) = \rho_0 + a*T$. The first and the second terms represent residual resistivity and the electron-phonon interaction terms, respectively. $\rho(T)$ behavior of this kind has been observed earlier for half-metallic NiMnSb and PtMnSb at low temperatures (up to nearly 20 K).[32] In the case of CFCA, $\rho(T)$ follows a linear behavior up to relatively higher temperatures (up to 150 K), indicating half-metallicity at temperatures as high as 150 K. In the region 2, (150-300 K), $\rho(T)$ is fitted well with the relation $\rho(T) = \rho_0 + b*T^n$, with n = 0.2. The $T^2$ term arising due to the electron-magnon interaction is found to be absent for $\rho(T)$ up to 300 K. The absence of electron-magnon interaction term in the electrical resistivity data indicates the absence of minority carriers at the Fermi level. This is in agreement with the high value of current spin polarization mentioned above. The *ab-initio* calculations which find a large half-metallic band gap of 0.31 eV also support the inference derived from the temperature dependence of resistivity data. The magneto-resistance (MR) is found to be non-saturating with field, with a value of -1.6% at 5 K for 50 kOe.

In conclusion, bulk CFCA alloy was investigated by x-ray diffraction, magnetization, spin polarization and magneto-transport measurements. The *ab-initio* electronic structure calculations predict half metallic nature for the alloy. The alloy is found to exist in B2 type cubic Heusler structure. Saturation magnetization of ~ 2.0 $\mu_B$/f.u. was found at 8 K, which is in good agreement with the value predicted by the Slater-Pauling rule (2.0 $\mu_B$/f.u.), $M_S$ value was found to be constant with pressure. The current spin polarization, $P = 0.67 \pm 0.02$ was deduced from PCAR measurements. Electrical resistivity followed a linear behavior with temperature up to relatively higher temperatures, indicating the possibility of high spin polarization at elevated temperatures. Considering the high value of P and electrical resistivity behavior, this material seems to be quite promising for the spintronic devices.

One of the authors, LB would like to thank UGC, Government of India for providing the senior research fellowship. AIM acknowledges the support from TAP fellowship under SEED Grant (project code 13IRCCSG020). KGS thanks ISRO, Govt. of India for the funds for carrying out this work. Authors thanks D. Buddhikot, TIFR for his help in magneto-transport measurements.

[1] K. Inomata, N. Ikeda, N. Tezuka, R. Goto, S. Sugimoto, M. Wojcik and E. Jedryka, Sci. Technol. Adv. Mater. **9**, 014101 (2008).

[2] B.S.D.Ch.S. Varaprasad, A. Rajanikanth, Y.K. Takahashi and K. Hono, Acta Materialia **57**, 2702 (2009).

[3] B.S.D.Ch.S. Varaprasad, A. Srinivasan, Y.K. Takahashi, M. Hayashi, A. Rajanikanth, K. Hono, Acta Materialia **60**, 625 (2012).

[4] L. Bainsla, K. G. Suresh, A. K. Nigam, M. Manivel Raja, B.S.D.Ch.S. Varaprasad, Y. K. Takahashi, and K. Hono, J. Appl. Phys. **116**, 203902 (2014).

[5]T. M. Nakatani, A. Rajanikanth, Z. Gercsi, Y. K. Takahashi, K. Inomata and K. Hono, J. Appl. Phys. **102**, 033916 (2007).

[6]A. Rajanikanth, Y. K. Takahashi and K. Hono, J. Appl. Phys. **101**, 023901 (2007).

[7]Y. Sakuraba, M. Ueda, Y. Miura, K. Sato, S. Bosu, K. Saito, M. Shirai, T. J. Konno, and K. Takanashi, Appl. Phys. Lett. **101**, 252408 (2012).

[8]Y. Du, B. S. D. Ch. S. Varaprasad, Y. K. Takahashi, T. Furubayashi and K. Hono, Appl. Phys. Lett. **103**, 202401 (2013).

[9]T. Kubota, S. Tsunegi, M. Oogane, S. Mizukami, T. Miyazaki, H. Naganuma, and Y. Ando, Appl. Phys. Lett. **94** (2009) 122504.

[10]Ikhtiar, S. Kasai, A. Itoh, Y. K. Takahashi, T. Ohkubo, S. Mitani and K. Hono, J. Appl. Phys. **115**, 173912 (2014).

[11]T. Saito, N. Tezuka, M. Matsuura and S. Sugimoto, Appl. Phys. Exp. **6**, 103006 (2013).

[12]K. Özdogan, E Sasioglu and I. Galanakis, J. Appl. Phys. **113**, 193903 (2013); X. Dai, G. Liu, G. H. Fecher, C. Felser, Y. Li, and H. Liu, J. Appl. Phys. **105**, 07E901 (2009).

[13]L. Bainsla, A. I. Mallick, M. Manivel Raja, A. K. Nigam, B.S.D.Ch.S. Varaprasad, Y. K. Takahashi, A. Alam, K. G. Suresh, K. Hono, arXiv:1410.0177.

[14]V. Alijani, S. Ouardi, G. H. Fecher, J. Winterlik, S. S. Naghavi, X. Kozina, G. Stryganyuk, and C. Felser, Phy. Rev. B **84**, 224416 (2011).

[15]P. Klaer, B. Balke, V. Alijani, J. Winterlik, G. H. Fecher, C. Felser, and H. J. Elmers, Phy. Rev. B **84**, 144413 (2011).

[16]L. Basit, G. H. Fecher, S. Chadov, B. Balke, and C. Felser, Eur. J. Inorg. Chem. **3950** (2011).

[17]H. S. Goripati, T. Furubayashi, Y. K. Takahashi, and K. Hono, J. Appl. Phys. **113**, 043901 (2013).

[18]H. Luo, H. Liu, X. Yu, Y. Li, W. Zhu, G. Wu, X. Zhu, C. Jiang, H. Xu, J. Magn. Magn. Mater. **321**, 1321 (2009).

[19]R. J. Soulen, J. M. Byers, M. S. Osofsky, B. Nadgorny, T. Ambrose, S. F. Cheng, P. R. Broussard, C. T. Tanaka, J. Nowak, J. S. Moodera, A. Barry, J. M. D. Coey, Science **25**, 282 (1998).

[20]G. J. Strijkers, Y. Ji, F. Y. Yang, C. L. Chien, J. M. Byers, Phys. Rev. B **63**, 104510 (2001).

[21]G. Kresse and J. Furthmuller, Phys. Rev. B **54**, 11169 (1996); Comput. Mater. Sci. **6**, 15 (1996).

[22]G. Kresse and D. Joubert, Phys. Rev. B **59**, 1758 (1999).

[23]H. J. Monkhorst and J. D. Pack, Phys. Rev. B **13**(12), 5188 (1976).

[24]G.Y. Gao, L. Hu, K.L. Yao, B. Luo, and N. Liu, J. Alloys Comp. **551**, 539 (2013).

[25]I. Galanakis, P. H. Dederichs, N. Papanikolaou, Phys. Rev. B **66**, 174429 (2002).

[26]F. J. Yang, Y. Sakuraba, S. Kokado, Y. Kota, A. Sakuma, and K. Takanashi, Phys. Rev. B **86**, 020409(R) (2012).

[27]S. Picozzi, A. Continenza, and A. J. Freeman, Phys. Rev. B **69**, 094423 (2004).

[28]T. Kanomata, Y. Chieda, K. Endo, H. Okada, M. Nagasako, K. Kobayashi, R. Kainuma, R. Y. Umetsu, H. Takahashi, Y. Furutani, H. Nishihara, K. Abe, Y. Miura, M. Shirai, Phys. Rev. B **82**, 144415 (2010).

[29]S. Picozzi, A. Continenza, and A. J. Freeman, Phys. Rev. B **66**, 094421 (2002).

[30]I. I. Mazin, Phys. Rev. Lett. **83**(7), 1427 (1999).

[31]S. K. Clowes, Y. Miyoshi, O. Johannson, B. J. Hickey, C. H. Marrows, M. Blamire, M. R. Branford, Y. V. Bugoslavsky, L. F. Cohen, J. Magn. Magn. Mater. **272**, 1471 (2004).

[32]J. S. Moodera and D. M. Mootoo, J. Appl. Phys. **76**, 6101 (1994).

**Figures**

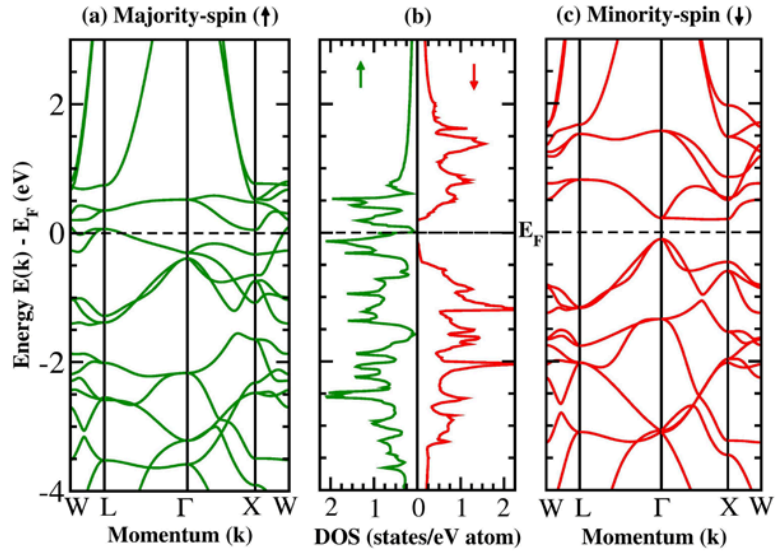

FIG. 1. Band structure and density of states of CFCA: (a) majority-spin bands, (b) density of states, (c) minority-spin bands at experimental lattice parameter ($a_{expt}$ = 5.75 Å).

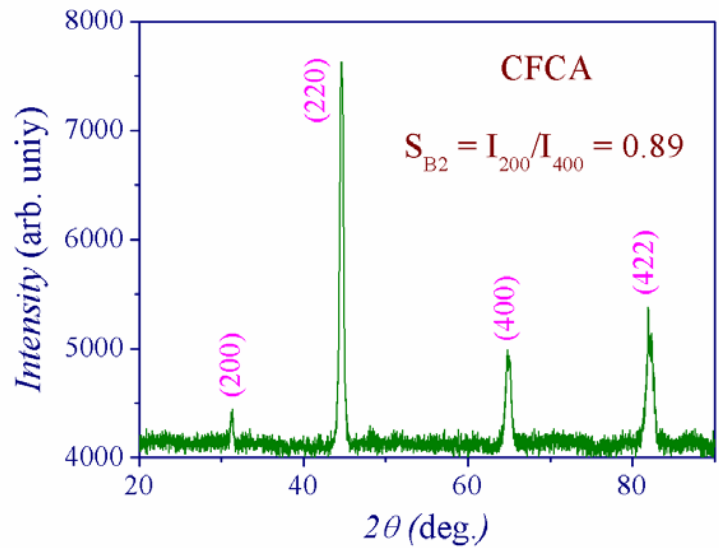

FIG. 2. Powder x-ray diffraction pattern of CFCA at room temperature.

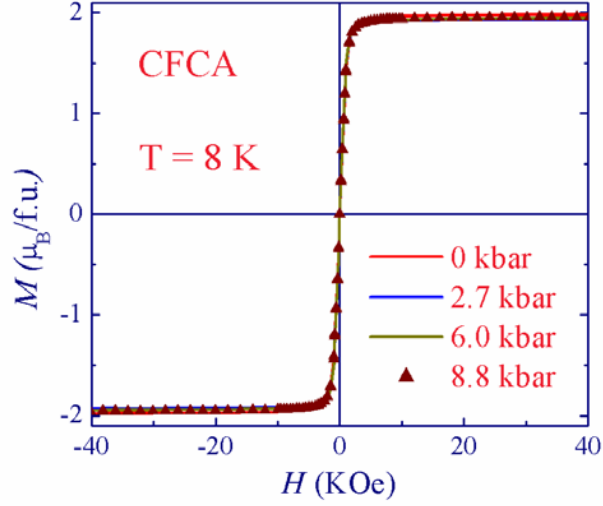

FIG. 3. Isothermal magnetization curves of CFCA obtained under various applied hydrostatic pressures at 8 K.

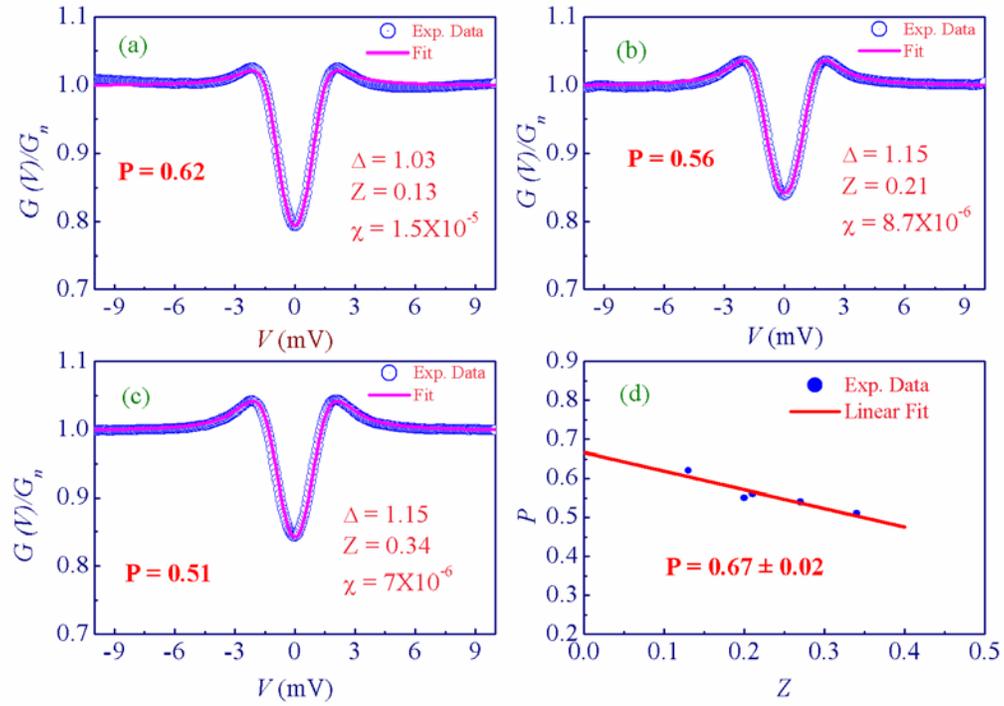

FIG. 4. (a) – (c) Normalized conductance curves of CFCA obtained at 4.2 K. The open circles denote the experimental data and the solid lines are the fit to the data using modified BTK model. (d) P *vs.* Z curve with the linear fit.

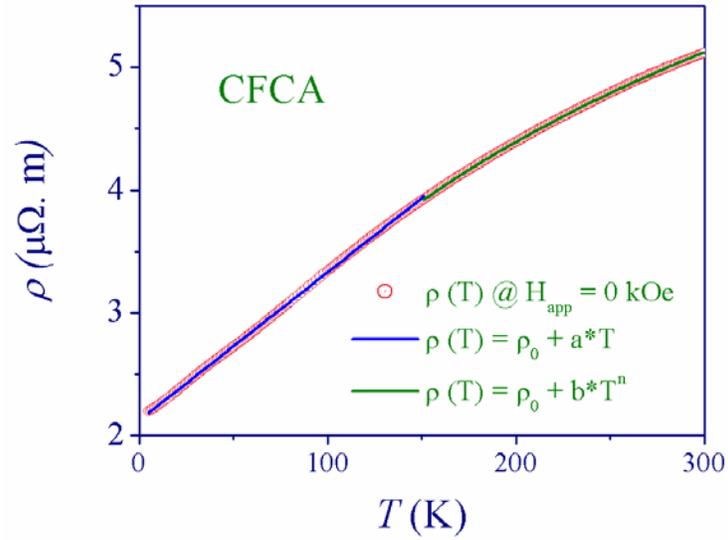

FIG. 5. Electrical resistivity variation with temperature of CFCA in the range of 5-300 K. The open circles denote the experimental data and the solid lines are the fit to the data to the equation $\rho(T) = \rho_0 + a*T$ in the low temperature regime (5-150 K) and $\rho(T) = \rho_0 + b*T^n$ in the range of 150-300 K.